\documentclass{emulateapj}
\usepackage{color}
\usepackage{natbib}

\newcommand{\mass}{\mathcal{M}}

\newcommand{\degree}{\ensuremath{^\circ}}

\newcommand{\Kepler}{\textit{Kepler}}
\newcommand{\EPOXI}{\textit{EPOXI}}
\newcommand{\Spitzer}{\textit{Spitzer}}

\shorttitle{Transit Timing Variations}
\shortauthors{Meschiari et al.}

\begin{document}

\title{Systemic: A Testbed for Characterizing the Detection of Extrasolar Planets. II. {Numerical approaches} to the Transit Timing Inverse Problem.}

\author{Stefano Meschiari \& Gregory P. Laughlin}

\affil{UCO/Lick Observatory, 
Department of Astronomy \& Astrophysics, 
University of California at Santa Cruz,
Santa Cruz, CA 95064}

\email{smeschia@ucolick.org}

\begin{abstract}
Transit timing variations -- deviations from strict periodicity between successive passages of a transiting planet -- can be used to probe the structure and dynamics of multiple-planet systems. In this paper, we examine prospects for {numerically solving} the so-called inverse problem, the determination of the orbital elements of a perturbing body from the transit timing variations it induces. We assume that the planetary systems under examination have a limited number of Doppler velocity measurements, and show that {a more extensive radial velocity characterization with precision comparable to the semiamplitude of the perturber may remove degeneracies in the solution}. We examine several configurations of interest, including (1) a prototypical non-resonant {system}, modeled after HD40307 b and c, which contains multiple super-Earth mass planets, (2) {a hypothetical system} containing a transiting giant planet with a terrestrial-mass companion trapped in low-order mean motion resonance, and (3) {the HAT-P-13 system}, in which forced precession by an outer perturbing body that is well characterized by Doppler radial velocity measurements can give insight into the interior structure of a perturbing planet, and for which the determination of mutual inclination between the transiting planet and its perturber is a key issue.
\end{abstract}


\section{Introduction}
{While the overall census of extrasolar planets continues to climb steadily (453 as of this writing\footnote{{exoplanet.eu, retrieved on May 12, 2010}}),  the emerging population of Earth and Super-Earth sized  ($\mass \sin i \leq 10 \mass_{\earth}$) planetary companions that has been uncovered by high-precision radial velocity (RV) surveys \citep[e.g.][]{Rivera05, Udry07, Mayor09, Vogt10} is shifting the interest of many planet search programs towards terrestrial planets}. Future refinements in ground-based RV programs will likely continue to further push the detection capabilities towards the low-mass end of the planetary population \citep{Mayor08, Howard10}.

On the other hand, the availability of ground and space-based surveys dedicated to photometric monitoring of large samples of host stars is affording constraints on the true mass and bulk composition of the Super-Earth planetary population \citep{Leger09, Queloz09, Charbonneau09}.  In particular, the \Kepler{} mission \citep[e.g.][]{Koch04, Koch10} is expected to yield transiting Earth-mass planets in the Habitable Zone (HZ) as part of its mission objectives, through continuous and simultaneous photometric sampling of more than 100,000 dwarf stars.  However, this class of objects will likely represent a small percentage of the detections (given the constraints of the mission design), and a large number of Neptune-mass and giant planets will be detected as well \citep[e.g.][]{Borucki10, Borucki10_4b}.

The exquisite precision and sheer size of the \Kepler{} transit timing datasets of giant planets, as observed  during the projected four years to six year mission duration, opens up an alternative route to the detection of low-mass planetary companions. Indeed, transit timing variations (TTV) will be caused by gravitational perturbations exerted by additional planets, causing deviations from strictly periodic Keplerian orbits \citep{MiraldaEscude02, Holman05}. These can be used to infer the orbital elements of the perturbing planet \citep{Agol05}, or at least place limits on the presence of additional planets \citep[e.g.][]{Alonso08, MillerRicci08}. An approximate analytic estimate for TTV amplitude for a transiting planet and an external perturber is given by \citep{Holman05}
\begin{equation}
\delta t \approx \ \frac{45\pi}{16}\ \left(\frac{\mass_{pert}}{\mass_*}\right) \ P_{trans}\ \alpha_e^3 \ \left(1-\sqrt{2} \alpha_e^{3/2}\right)^{-2}\, 
\end{equation}\label{eqn:ttv}
(where we use the symbols $\mass$ for mass, $P$ for period, $e$ for eccentricity and $a$ for semi-major axis; $\alpha_e = a_{trans} / \left[ a_{pert} (1-e_{pert})\right]$).

The amplitude of these variations can be quite large and amenable to detection, {either in the presence of high-eccentricity perturbers \citep[e.g.][]{Steffen05}} or when the two planets lie near a low-order mean motion resonance (MMR). Indeed, MMRs are an entirely plausible outcome of core-accretion models of planetary formation, whereby planets can be captured and locked into an MMR during the migration stage \citep[e.g.][]{Nelson02, Papaloizou05, Beauge06}. Observationally, several of the detected extrasolar systems with multiple planets may be locked in low-order MMRs. {Three such systems (HD82943, HD73526 and HD128311) are engaging in deep 2:1 resonances well characterized by the observations, and GJ876 has recently been reported as a Laplace-type resonance chain \citep[4:2:1;\ ][]{Rivera2010}}. {For instance, the TTV amplitude induced by an Earth-mass perturber in a 2:1 resonance with a 3-day Jupiter-mass planet, both in circular orbits, is of order of minutes \citep{Agol05}}. This is a large signal compared to an accuracy in the measurement of the central transit time of order \citep{Ford06_2}
\begin{equation}
\sigma_T \approx \left(\frac{t_e}{2\Gamma}\right)^{1/2} \sigma_{ph} \left(\frac{R_{pl}}{R_*}\right)^{-2}
\end{equation}
{(where $t_e$ is the duration of the transit ingress/egress, $\Gamma$ is the observation rate, $\sigma_{ph}$ is the photometric precision, $R_{pl}$  and $R_*$ are the radius of the planet and the radius of the star, respectively)}, amounting to 10s of seconds for milli-mag photometric accuracy. The recently published \Kepler{} central times \citep{Latham10, Borucki10_4b, Jenkins10, Dunham10, Koch10} are in rough accordance with this estimate. Furthermore, with respect to the \Kepler{} project, we note that once a transit is detected with sufficient signal-to-noise ratio, the star will be switched from the long-cadence (30 minute) to short-cadence (1 minute) sampling rate \citep{Borucki08}, improving the temporal resolution of the transit even further. We take $\sigma_{tr, K} = 2\times 10^{-4}$ d ($\approx$ 15 s) as a conservative estimate of accuracy on the central transits. 

Given a large dataset comprising 1 year or more of continuous transit monitoring, is it possible to infer the mass and elements of the perturbing planet? Reconstructing the properties of the perturber from a noisy TTV signal is a complex, and possibly highly degenerate \citep{Nesvorny08}, inverse problem. In this paper, we present a series of simulations aimed at detecting low-mass perturbers from realistic central transit and follow-up RV data. To this end, we produce a large sample of \Kepler-like observations and attempt to characterize the perturber using the algorithm toolset offered by a revised version of the Systemic Console \citep[][hereafter Paper I]{Meschiari09}.  A number of different planetary realizations were used, in an attempt to fully capture the complexity of TTV fitting, drawing the orbital elements from observed planetary systems. For the sake of simplicity, we focus on two-planet systems, but the method is fully general within the constraints of CPU time and measurement errors.

The plan of the paper is as follows. In \S \ref{sec:Console}, we briefly review describe the algorithms used to derive best-fit models and accompanying error estimates. In \S \ref{sec:HD40307}, we examine the characterization of planets similar to HD40307\emph{c} and \emph{d}, which lie close but not quite in a 2:1 MMR. Our analysis makes use of the HARPS dataset  \citep{Mayor09} and a simulated transit timing dataset. In \S \ref{sec:HATP7} we fit the synthetic realization of a planetary system deep in a 2:1 MMR, with an external perturber (using HAT-P-7 as our model system).  Finally, in \S \ref{sec:HATP13} we analyze constraints placed by TTVs on the three-dimensional configuration of planetary systems, using HAT-P-13 as a test case, and conclude in \S \ref{sec:conc}.

\section{Numerical setup}\label{sec:Console}
The transit timing variation signal is defined as the difference between the observed central transit times and the predicted times from a linear regression (corresponding to a single-planet Keplerian fit with period $P_1$):
\begin{equation}
\delta t_k  = t_k - k P_1
\end{equation}
The variations originate by the mutual gravitational interactions with an additional body, chiefly causing short-term oscillations wherein the true anomaly $f_1$ trails or leads the Keplerian value and long-term effects such as pericenter precession \citep{Heyl07}. In principle, since the signal will depend on the Newtonian evolution of the planetary system, TTVs can provide a sensitive probe for the three-dimensional orbit of the second planet, in combination with the tight constraints on the eclipsing planet's period and the time of pericenter passage provided by the central transits themselves.

However, solving the inverse problem of deriving a best-model fit to the TTV observations can be daunting. The computation of the predicted TTV signal requires precise N-body integrations (with $N \geq 3$). In the general case, the dependence of the signal on the set of orbital parameters is not directly clear; unlike, e.g. the RV technique, deviations from the Keplerian signal -- as opposed to the Keplerian signal itself -- constitute the bulk of the information. The use of Fourier analysis to sort out periodicities in the data is generally hampered by the sparseness of the transit observations. Furthermore, given the extreme sensitivity of $\delta t$ to the model parameters, local minimization routines can easily get stuck in narrow $\chi^2$ minima, or fail due to steep gradients in the landscape. Finally, as shown in the later sections, there is a degree of non-uniqueness as multiple models can fit the transit timing observations when measurement errors are taken into account \citep[see also e.g.][]{Nesvorny08}; these degenerate solutions are characterized by comparable $\chi^2\sim 1$, and must be taken into account when deriving parameter uncertainties.

Direct searches of the parameter space \citep[e.g.][]{Steffen07} can be extremely expensive in terms of CPU time. A more appealing alternative is represented by the TTV Inversion Method \citep[TTVIM; ][and related papers]{Nesvorny10}. TTVIM combines a fast  algorithm for computing the 2-planet transit timing based on perturbation methods with a downhill simplex method to obtain good convergence towards the perturbing planet's parameters. However, some issues remain in addressing systems lying close to a MMR. 

In this paper, we adopt the approach of finding best-fit models to joint TTV and Doppler velocity data sets by driving an efficient Bulirsch-Stoer integrator with the Simulated Annealing algorithm integrated in the Systemic Console (Paper I)\footnote{The new version of the Systemic Console, including a Bulirsch-Stoer integrator, AMOEBA and fully non-coplanar fitting is available for download at www.oklo.org}.  SA-type algorithms are well-suited to exploring the orbital parameter space {(period $P$, mass $\mass$, eccentricity $e$, inclination $i$, mean anomaly at epoch $M_0$, longitude of pericenter $\varpi$ and node $\Omega$ for each planet)} and converging, in principle, to global minima (subject to appropriate choices of scheduling algorithm and scale parameters). Several minimizers can be run in parallel with different initial temperatures and initial conditions, exploiting modern multi-core CPUs capabilities. The step size vector is automatically adjusted to attain an acceptance rate of $\sim 25$\%; we have found empirically that this value is an optimal compromise. After a fixed number of steps, we invoke a downhill simplex algorithm \citep[AMOEBA;][]{Press92} in an attempt to home in on nearby deep minima. This avoids missing promising solutions when the SA step size is too large to properly resolve them. In practice, this scheme permits the derivation of the full set of degenerate solutions compatible with the observational errors.

Although we recognize that this approach can be computationally inefficient compared to perturbation methods, the implementation is trivial and can use existing integration techniques. Furthermore, it permits the characterization of arbitrary planetary configurations (including $N_{pl}>2$, resonant, high-eccentricity and inclined bodies) and the inclusion of additional dynamics (such as tidal evolution) self-consistently, owing to the fully general N-body integration. Finally, we remark that in this work the parameters of the transiting planet are \emph{not} fixed, but derived simultaneously from the available data. This mimics follow-ups of transiting planets, whereby the mass of the transiting planet is determined by a small number of RV measurements.

We use the combined $\chi^2$ statistic detailed in Paper I to simultaneously fit the transit timing and follow-up RV datasets. While there is a degree of ambiguity in the choice of the weighing factor $\lambda$, this is not a concern in the vicinity of a solution, where the contribution from RVs and transits is approximately equal for $\lambda = 1$. Far from the solution, the contribution from transits to the $\chi^2$ budget is extremely large; however, this is not an issue in practice because we first fit for a one-planet solution, reducing the initial $\chi^2$ to $\chi^2 \sim \delta t / \sigma_{TR}$.

\section{HD40307}\label{sec:HD40307}
\citet{Mayor09} recently announced a three Super-Earth planetary system orbiting the nearby metal-deficient dwarf HD40307. Interestingly, while this system lies close to a 4:2:1 Laplace resonance chain, such a configuration is ruled out by the observations. The a-priori transit probability for the innermost 4.3d planet is high enough to warrant a transit follow-up; unfortunately, no transit was detected using \Spitzer{} \citep{Gillon10}, preventing the placement of desirable constraints on the bulk composition of  the three planets.
\begin{figure}
\plotone{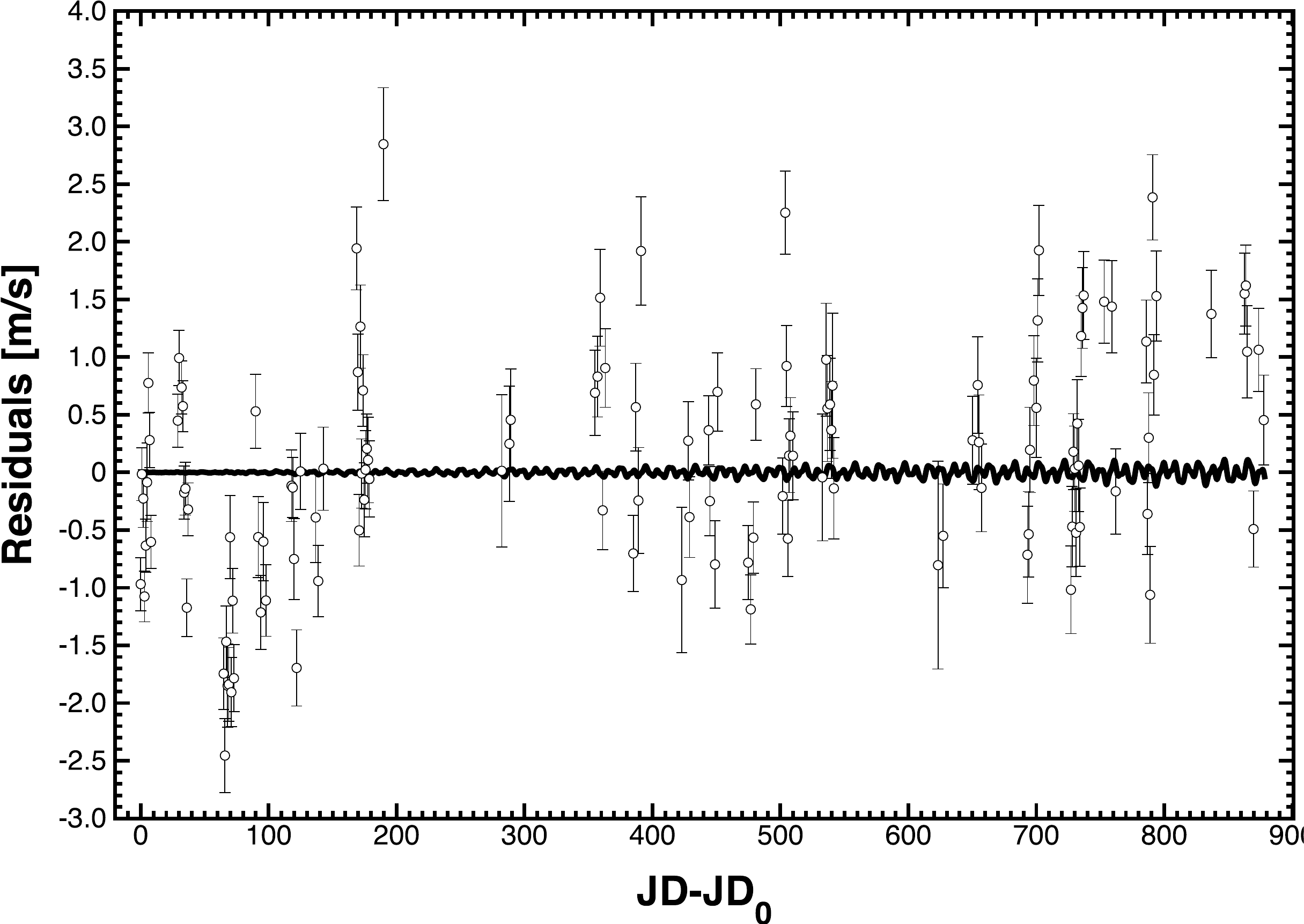}
\caption{Sensitivity of the RV method to the mutual gravitational perturbations: Keplerian model subtracted from the integrated model (thick curve) compared to the HARPS residuals (empty circles).}\label{fig:40307_rv}
\end{figure}

\begin{figure}
\plotone{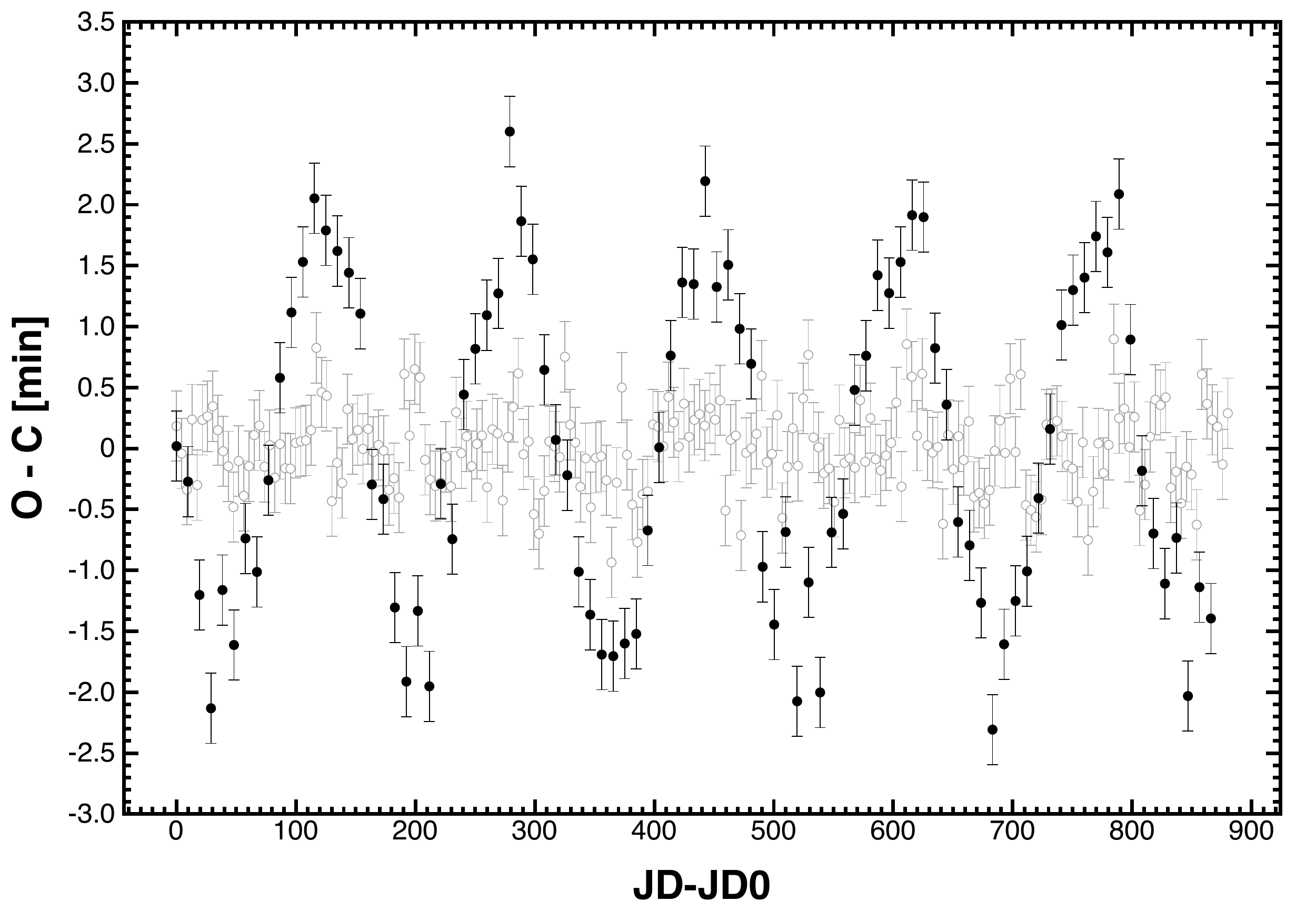}
\caption{Predicted transit timing variations for planets \emph{b} (empty circles) and \emph{c} (black circles), over the HARPS observation window.}\label{fig:40307_ttv}
\end{figure}

We derived the orbital parameters for the system using the publicly available HARPS dataset, obtaining best-fit and error estimations in good accordance with the published configuration.  As can be seen by Figure \ref{fig:40307_rv}, the difference in RV signal between a fully integrated model (using Bulirsch-Stoer) versus simple superposition of Keplerian orbits is negligible compared to the HARPS error bars and the RV residuals.

As a comparison, we derived synthetic TTV computed comparing a simple linear fit to synthetic transits computed with the fully integrated solution above; we assumed, respectively, planets \emph{b} and \emph{c} to be transiting and computed the primary transit timings for the HARPS observation window.  To each transit timing observation, we added a Gaussian white noise of amplitude $\sigma = 2\times 10^{-4}$ d (0.3 minutes), as a simple, conservative model for \Kepler{} timing uncertainties. The TTV dataset is shown in Figure \ref{fig:40307_ttv}. The amplitude of the TTV signal for planet \emph{c} is approximately 5$\sigma$, making it a far more sensitive probe of the mutual gravitational perturbations than the highest-precision RV measurements available.

Although no transits have been so far detected for planet $b$, the transit probability for planet \emph{c} is a tantalizing 5\% and the transit depth is of order 400 ppm, fully within the capabilities of \Kepler. Therefore, it is an interesting illustrative test-case problem to use the known orbital elements of the HD40307 system and analyze the constraints imposed by TTV on the perturbing planet, in absence of high-precision radial velocities. Given that the bulk of the signal originates from the mutual perturbation between planets \emph{c} and \emph{d}, we hereafter solve the simpler two-planet inverse problem and neglect the contribution from planet \emph{b}. The orbital elements of the generating fit are reported in Table \ref{tab:40307}.

\begin{deluxetable}{lccc}
\tablecaption{Best-fit solutions\label{tab:40307}}
\tablewidth{0pt}
\tablehead{\colhead{}&\colhead{Best fit (HARPS)}&\colhead{Best fit (100d)}&\colhead{Best fit (365d)}}
\startdata
\emph{P} (days)		& 9.621 (1)	&	9.6214 (4)	& 9.62114 (5)\\
						& 20.439 (5)&	20.2 (2)		& 20.45 (1)\\
$\mass$ ($\mass_J$)	& 0.0218 (6) &	0.021 (5)	& 0.02 (1)\\
						& 0.0290 (8) &	0.025 (4)	& 0.025 (2)\\
\emph{e} 				& 0.06 (3)	&	0.036 (3)	& 0.034 (4)\\
						& 0.12 (2) 	&	0.06 (3)		& 0.01 (2) \\
$\varpi$ ($^\circ$)		& 284 (6) 	&	358 (4)		& 358.2 (1)\\
						& 12 (7) 	&	78 (23)		& 71 (4)\\
\hline
$\chi^2$				& 10.49 	&	1.29		& 1.15 \\
RMS (m s$^{-1}$)		& 1.04 		& 	1.17		& 1.25 \\
$\chi^2_{TR}$			& -- 		&	0.4			& 0.75\\
\enddata
\tablecomments{Best-fit solutions for the HD40307 system.  The error on the least significant digit is indicated in parentheses.}
\end{deluxetable}

We generated two sets of central transit observations spanning 100 days (11 transits) and 365 (38 transits); we assumed every transit is detected with $\sigma_{tr} = 2\times 10^{-4}$ d. We also computed a small set of ``follow-up'' synthetic radial velocity observations (10 points), which set the scale for the mass and the eccentricity of the transiting planet (period and mean anomaly at epoch being primarily determined by the transit timing). We draw the measurement errors to mimic mid-range precision observations; the average measurement error is $\sim 1.5$ m/s. As a comparison, we also computed a third synthetic RV dataset drawing from the HARPS schedule and measurement errors for the two planets alone; we assumed a small jitter of $\sim 0.7$ m/s. We point out that this jitter is excellent, and depending on the properties of the parent star, a realistic case might require more RVs.

Eight SA simulations were launched (one per core on a Mac Pro Xeon workstation), with initial temperature and step size regulated to achieve 25\% acceptance rate in each orbital parameter. 
{The initial configuration used the parameters from a single planet best-fit for the transiting planet, and random elements for the perturbing planet (period, mass, eccentricity, mean anomaly and longitude of pericenter), avoiding orbit-crossing configurations. For the sake of efficiency, we constrained the period of the second planet between 1.5 and 5 times the period of the inner planet, the masses between 0.3 and 32 $\mass_\earth$ and the eccentricity between circular and 0.5.} This parameter range approximately spans the region where the transit timings are sensitive to the perturbations, but the reflex stellar semiamplitude $K$ is not so large to be readily picked up by RV observations. Finally, every 2,000 steps the current configuration was submitted to the AMOEBA routine to attempt direct convergence to a solution. The minimization routine is considered to be converged and the solution is retained if $\chi^2_{TR} < 1.1$ and the radial velocity $RMS < \overline{\sigma}_{RV}$, corresponding to a combined $\chi^2 \sim 1.3$. After a predetermined number of steps (10,000), if no improvement in the total $\chi^2$ has been reached, the suboptimal solution is discarded and a new set of initial conditions is chosen. A sample of 20 candidate solutions was derived for each dataset; each solution representing a local minimum. The lowest $\chi^2$ solution was chosen as the representative best-fit.
\begin{figure}
\plotone{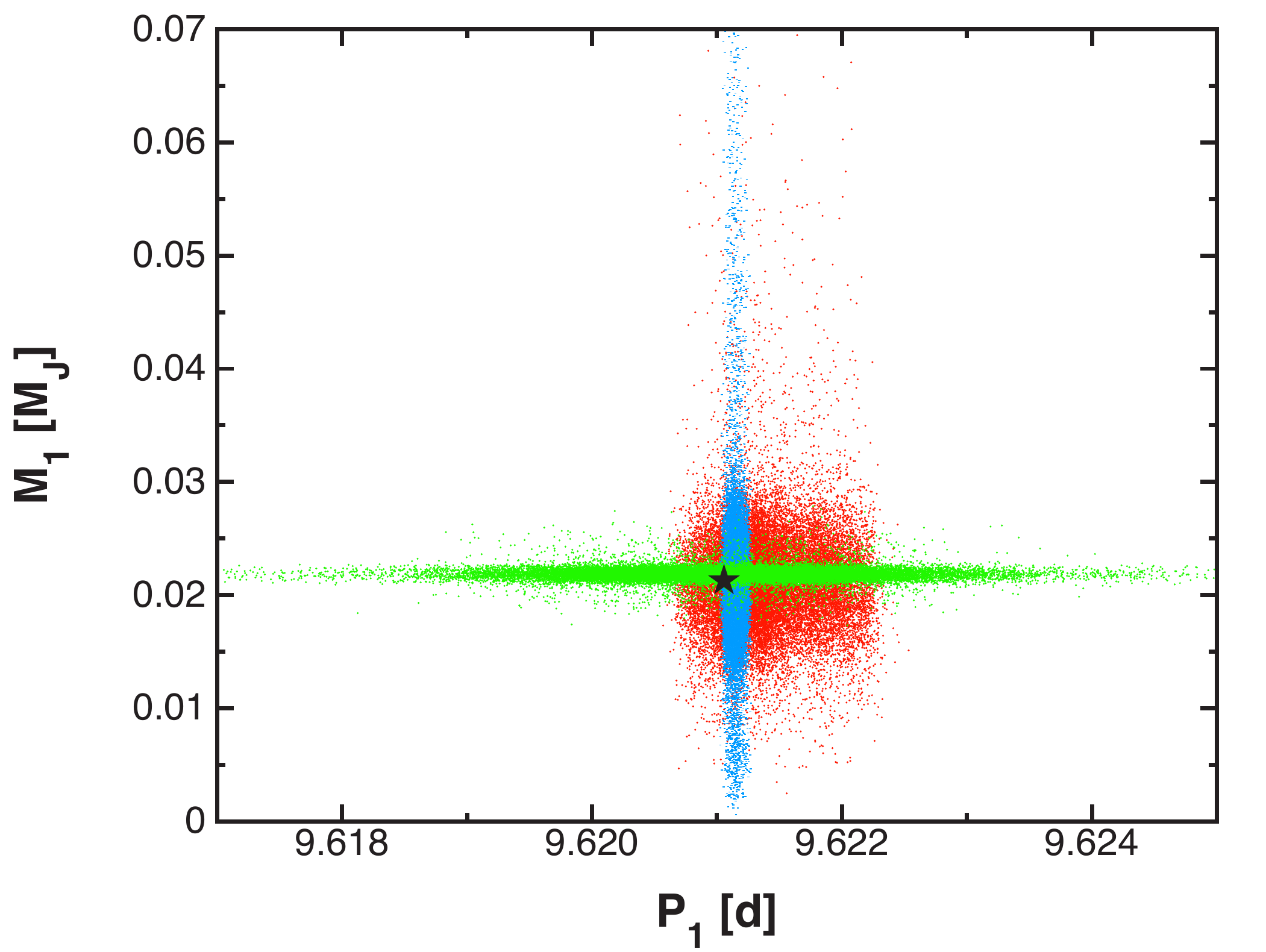}\\
\plotone{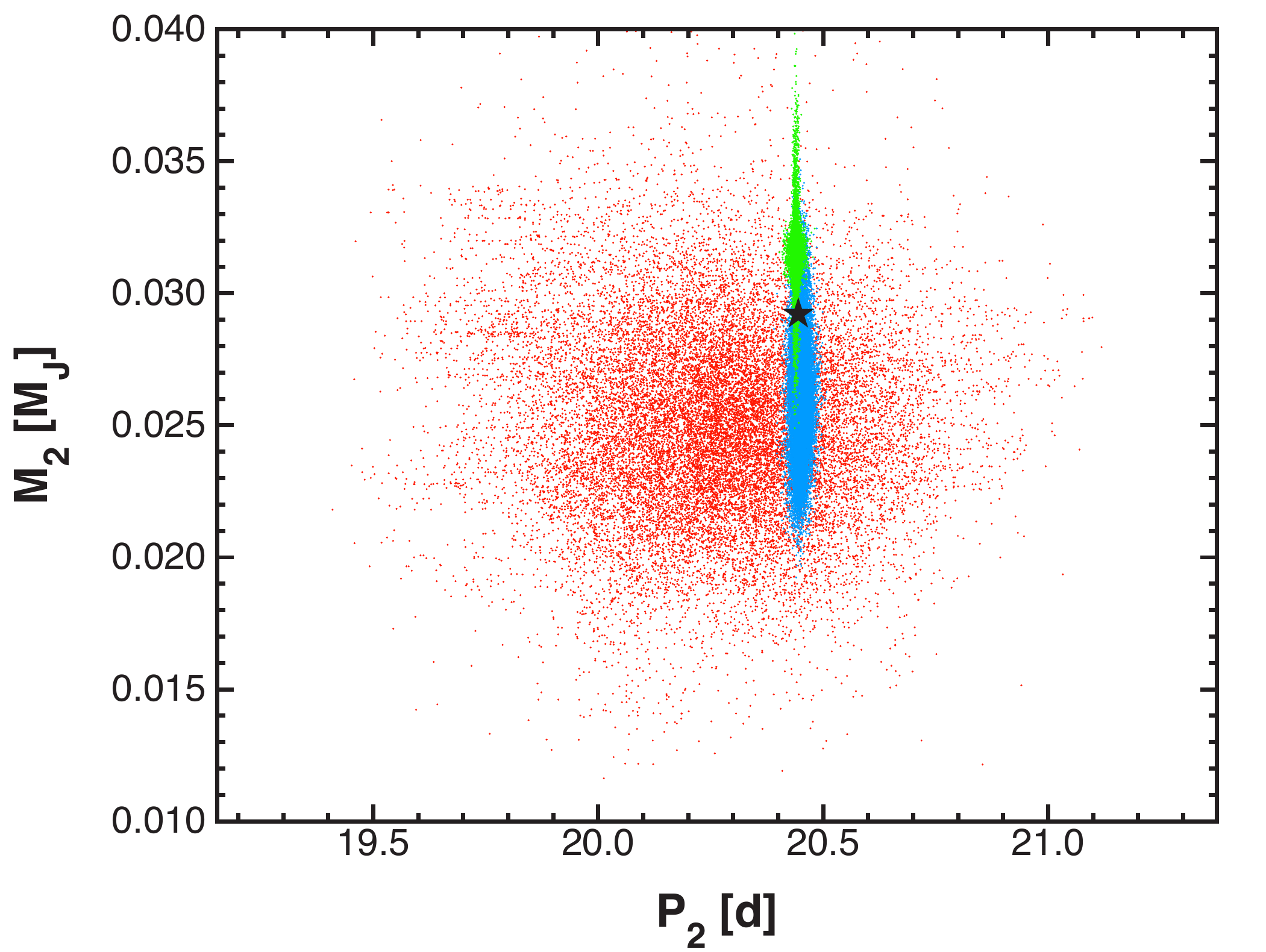}\\
\caption{Results of a MCMC simulation consisting of 50,000 states computed from the HARPS dataset (green points), 100 days of transit timing observations + 10 follow-up RVs (red points), 365 days of transit observations + 10 follow-up RVs (blue points). The parameters of the originating system are marked with a star symbol.}\label{fig:40307_mcmc}
\end{figure}

An estimation of the uncertainties on the orbital parameters of the best-fit solutions was derived using the Markov Chain Monte Carlo algorithm \citep[MCMC;][]{Ford05}. While the synthetic HARPS dataset is amenable to a bootstrap resampling technique, the rugged $\chi^2$ landscape for the TTV dataset turned out to be excessively complicated for an efficient exploration, yielding artificially low parameter uncertainties. The simple MCMC algorithm presented in Paper I derived error bars in accordance with bootstrap estimates for the RV dataset. We use uniform priors in \{$\log P$, $\log \mass$, $M_0$, $e$, $\varpi$\}; while 
more sophisticated approaches are available (e.g. incorporating information from Eqn. \ref{eqn:ttv} as a constraints), the size and precision of the synthetic datasets provide strong constraints on the model parameters and the choice of the priors should not affect our results \citep{Ford06}. We construct MCMC chains of 50,000 states, each state consisting of 200 iterations. The initial 10\% portion of the chain is considered ``burn-in'' and discarded. 

The best-fit solutions to the three datasets (synthetic HARPS, 100-d and 365d TTVs + RV followup) and respective uncertainties are compared in Table \ref{tab:40307}. To our knowledge, this is the first attempt to fully fit and derive error estimates on a large TTV dataset. We show the parameter scatter for the second planet in Figure \ref{fig:40307_mcmc}.

The computed parameter uncertainties show a number of interesting properties. Firstly, the period and mass of the second planet are derived to an accuracy comparable to that of the full HARPS dataset, which spans 4.5 years. The detection of a low-mass planet at this level of accuracy showcases the potential of scanning the future \Kepler{} datasets for TTV detection candidates. Once again, we stress that our estimate of the central transit timing noise is likely conservative and that stars on the short-cadence list will be observed with an even higher accuracy. While the period of the transiting planet is constrained by the transits timing themselves, the mass is not well constrained because, to a good approximation, the amplitude of the TTVs does not depend on the mass of the transiting planet itself (Equation \ref{eqn:ttv}) {in the non-resonant regime}. 

Finally, we remark that although the $\chi^2$ landscape allowed for several, well-separated local minima, both the SA and the MCMC algorithms were able to efficiently sample the parameter space. Therefore, it is likely that global minimization routines will be part of the standard toolset to analyze the future \Kepler\ transit datasets.

\section{HAT-P-7}\label{sec:HATP7}
\begin{figure}
\plotone{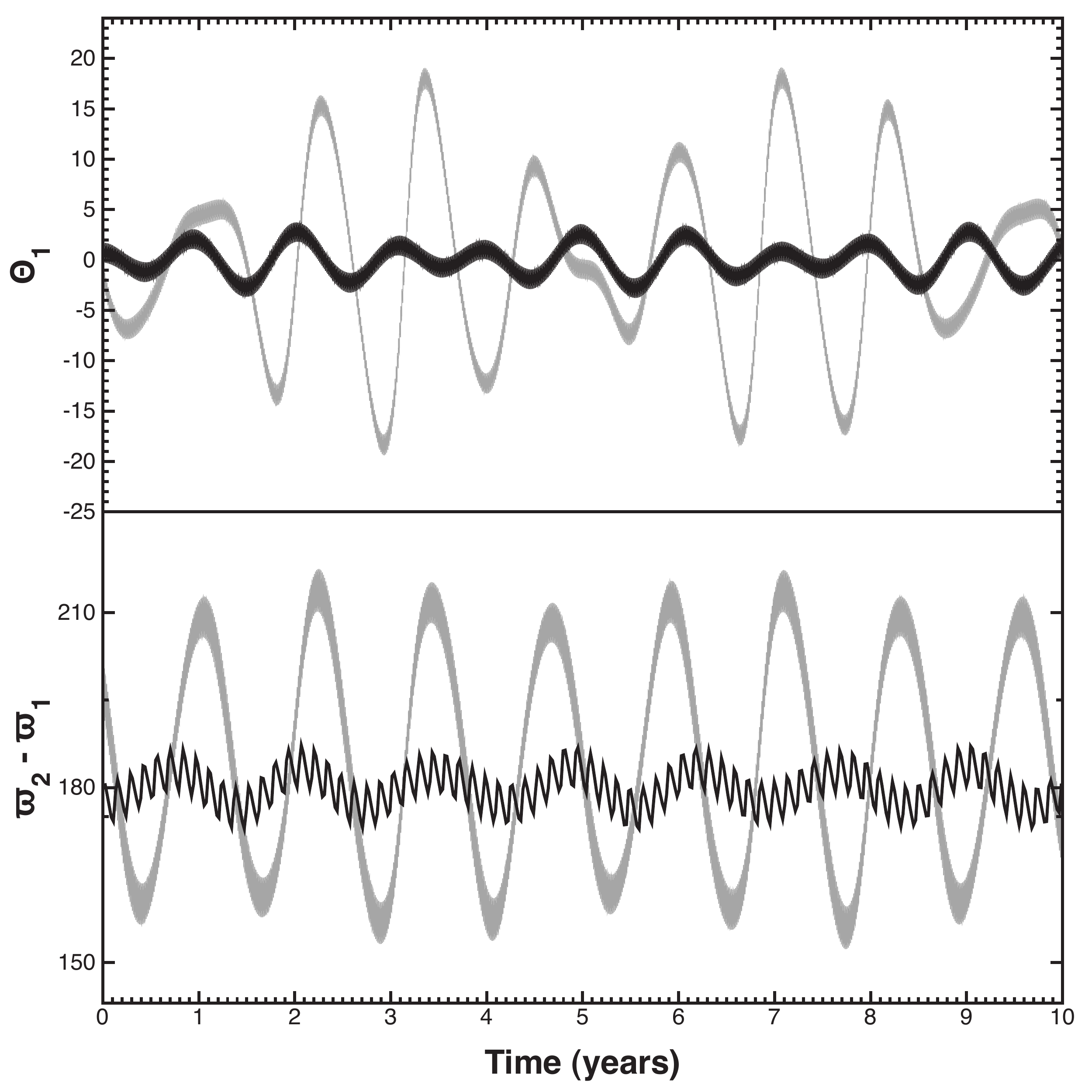}
\caption{Libration (in degrees) of the resonant arguments $\Theta_1$ and $\Delta \varpi$ for the reference configuration (black line) and the best-fit configuration (grey line).}\label{fig:resonance}
\end{figure}

\begin{figure}
\plotone{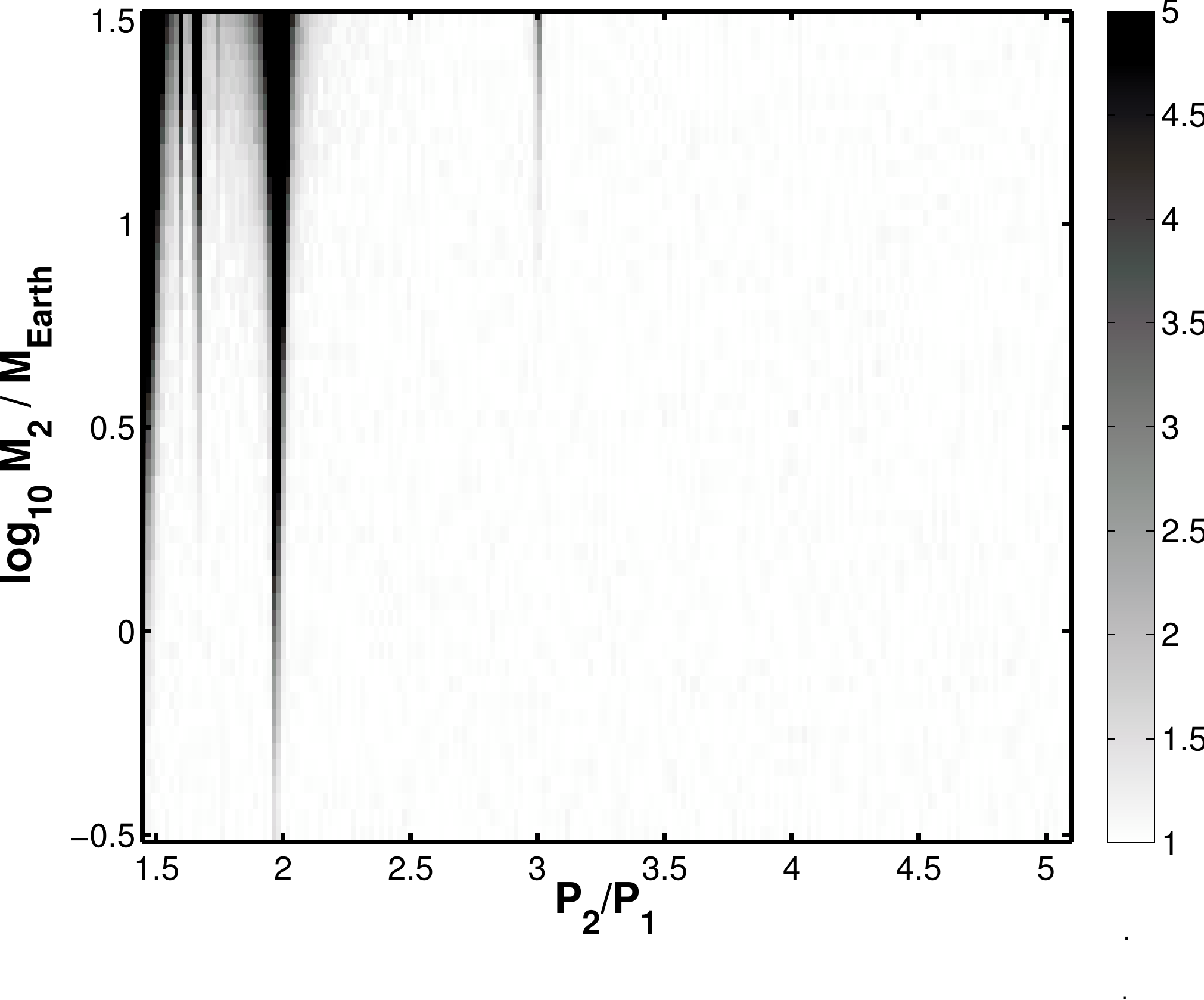}
\plotone{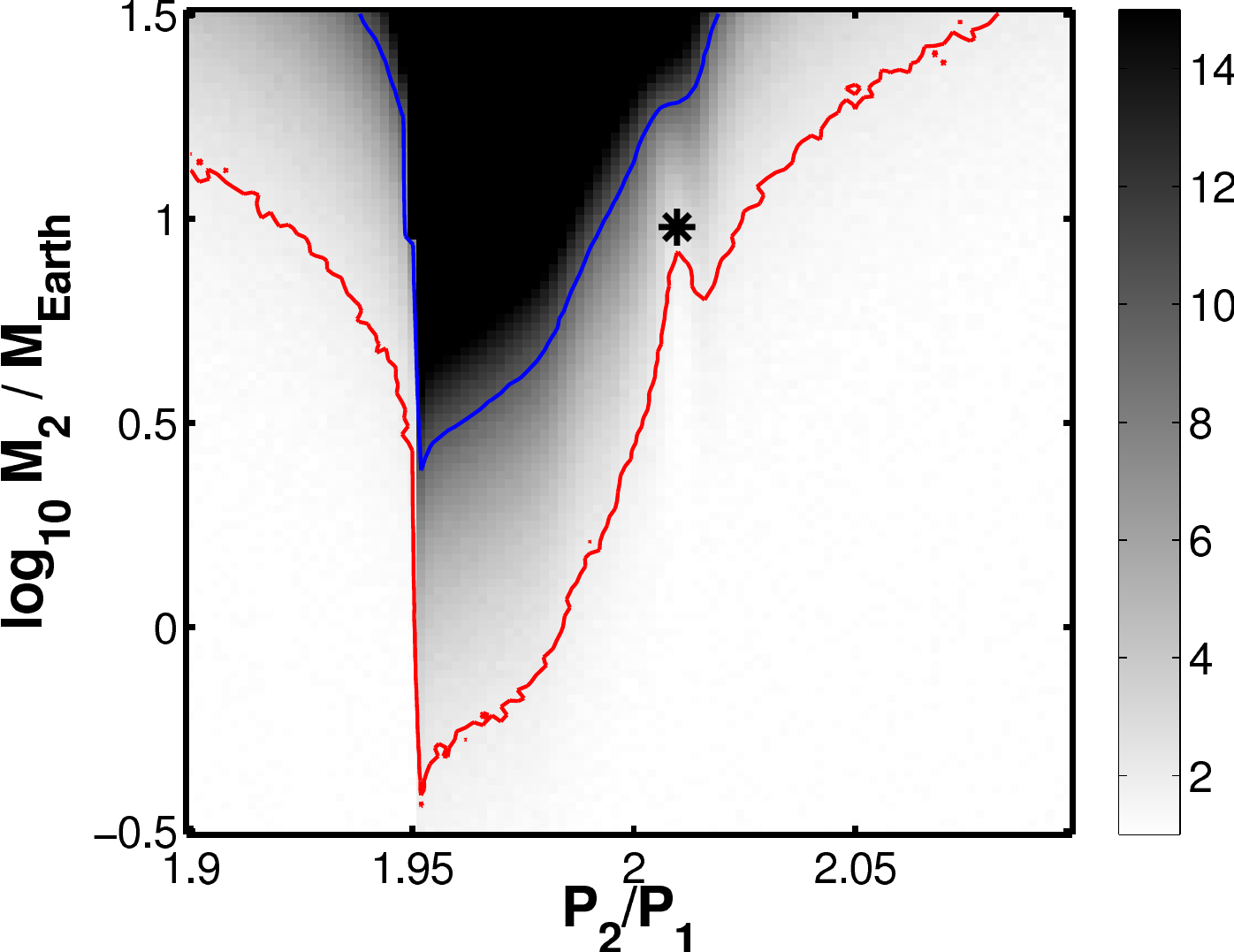}
\caption{\emph{(Top)} Grayscale map of $\delta t$ (in units of $\sigma_{tr, K} \approx 15$ s) for 10,000 realizations spanning a range of perturber periods and masses, using the reference configuration for the other elements. \emph{(Below)} As above, in the region near the 2:1 resonance.  The contours show the parameter space where $\delta t > 2\sigma_{tr}$, assuming $\sigma_{tr} = 2\times 10^{-4}$ d (\Kepler, red contour) and $\sigma_{tr} = 10^{-3}$ d (\EPOXI{}, blue contour) respectively. The star symbol represents the reference configuration.} \label{fig:map}
\end{figure}

\begin{figure}
\plotone{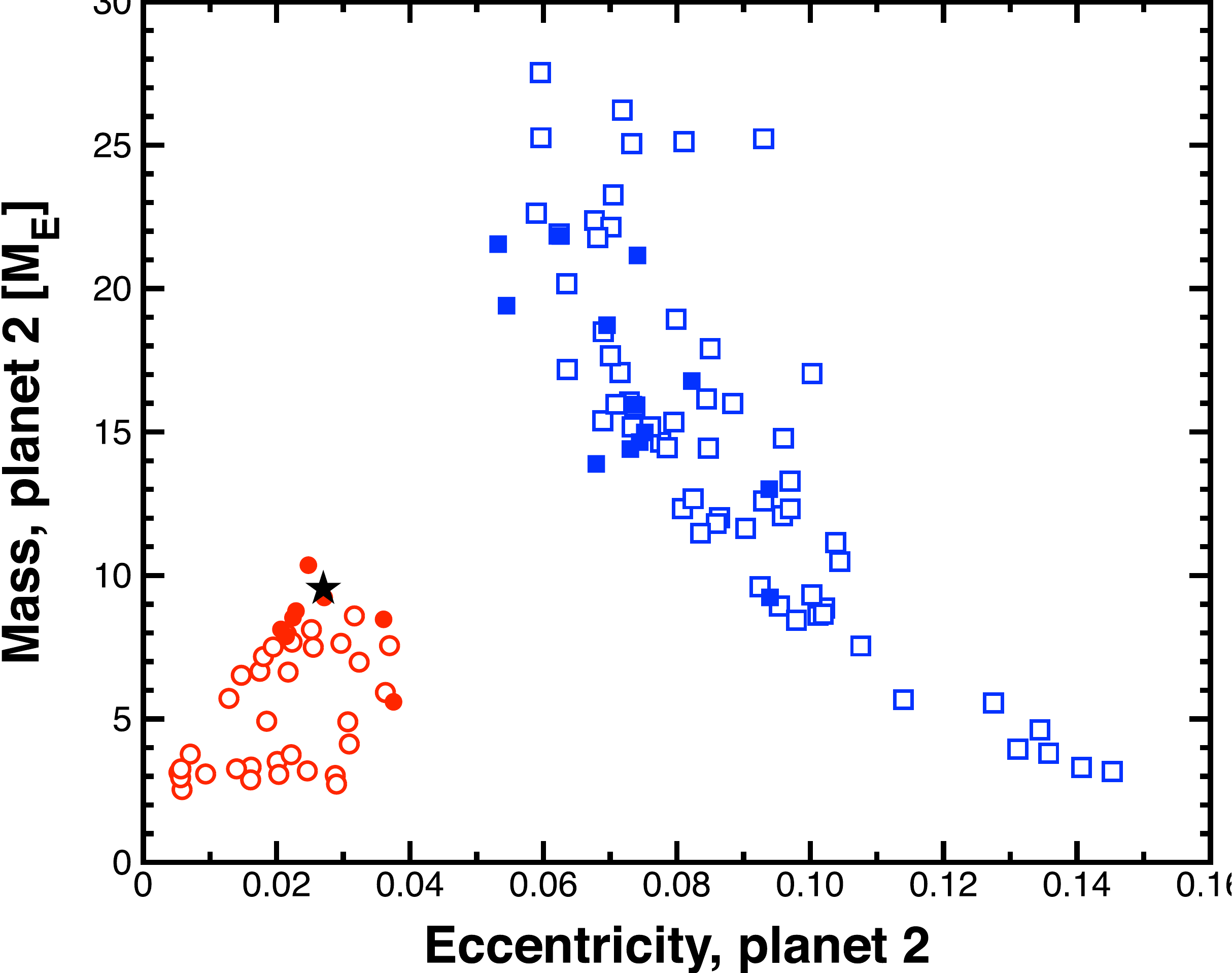}
\caption{Best-fit solutions for the HAT-P-7 dataset lying near the 2:1 resonance (circles) and the 3:1 resonance (squares), for two different levels of noise in the TTV measurements: $2\times 10^{-4}$ (empty symbols) and $5\times 10^{-5}$ (filled symbols).}\label{fig:hatp7sols}
\end{figure}

The bright  nearby dwarf HAT-P-7 hosts a transiting hot Jupiter, first characterized by the HATNet project \citep{Pal08}. The star is in the field of view of one of the \Kepler\ detectors; ten days of photometric data, as processed by the \Kepler{} pipeline, were obtained during the commissioning phase \citep{Borucki09_2}. Additional primary transits and a number of secondary eclipses were observed using \EPOXI{} and \Spitzer{} \citep{Christiansen10}, with the intent of studying the atmospheric properties of the planet. The EPOXI best-fit central times achieved an accuracy of $\sigma_{tr} \approx 10^{-3}$ d ($\approx 1.5$ minutes).

Given its extensive and diverse coverage, and the inclusion of this planet in the \Kepler{} star list, we chose this system as a prototype of the class of massive transiting planets that will be monitored by the \Kepler{} mission and may reveal TTVs. In particular, we are interested in assessing the secure detection of a low-mass planet in a 2:1 MMR with the transiting gas giant (we consider only the case of an external perturber in the present analysis). 

We generated a realistic resonant configuration self-consistently with the following procedure. We placed the two planets {(denoted as 1 and 2, respectively the transiting planet and the external perturber)} on originally widely separated orbits; following \citet{Lee02}, we added a forced migration ($\dot{a}/a = -3\times 10^{-4}\ \mbox{yr}^{-1}$) and an eccentricity damping ($\dot{e}/e = 100\ \dot{a}/{a}$) term of the outer planet to the equations of motion until resonant capture is achieved. In this reference configuration, the outer planet was captured into an {antialigned} configuration with $\Theta_1 = 2\lambda_2 - \lambda_1 - \varpi_2$ librating around 0\degr\ and $\Delta\varpi = \varpi_2 - \varpi_1$ librating around 180\degr, with an amplitude of $\approx 5$\degr (Figure \ref{fig:resonance}).  {The final eccentricities for this choice of forced migration terms are low ($e_1 = 0.002$, $e_2$ = 0.027).}

{To illustrate the process, we chose a mass for the second planet of $\approx 10 M_{\earth}$, since this can yield a TTV signal larger than 1 minute, easily detectable with \Kepler. Figure \ref{fig:map} shows the amplitude of the TTV signal for a choice of periods and masses, at fixed eccentricies and phases; as expected, the TTVs are largest in the proximity of resonances. In particular, 3:2, 2:1 and 3:1 MMRs yield a sizable TTV signal for our range of perturber masses.}

We created a TTV dataset spanning 1 year (166 observations) following the procedure in Section \ref{sec:HD40307}, using the reference configuration as our generating system and Gaussian noise at the level of $2\times 10^{-4}$ d. We drew from the schedule and uncertainties of the Keck/HIRES follow-up observations \citep{Pal08} to generate the accompanying RV dataset. We note that given the small semi-amplitude $K_2$ ($\approx 2.8$ m/s, larger than the typical error in the Keck dataset but smaller than the stellar jitter $\approx 3.8$ m/s) and the few RV points available, the RV dataset places only a weak constraint on the parameters of the perturbing planet.

We launched a number of SA chains and allowed the parameters of the perturbing planet to float freely. We found that the best-fitting solutions comprised a set of degenerate configurations, shown in Figure \ref{fig:hatp7sols}.  The fitting routine found two groups of solutions: configurations lying near a 2:1 MMR and configurations lying near a 3:1 MMR can fit the TTV signal equally well. Additionally, the degeneracy between mass and eccentricity of the perturbing planet makes it impossible to place a strong constraint on the mass of the second planet. 

This non-uniqueness of the inverse problem was already noted in \citet{Nesvorny08}; the measurement errors filter out some of the TTV harmonics. {The authors also pointed out that the non-uniqueness threshold (the measurement uncertainty that leads to a unique solution)  of the number of transits detected}; accordingly, we verified that a transit dataset covering 2 years of observations still yielded the two groups of solutions. Reducing the error on the transit measurement to $5\times 10^{-5}$ d (4 seconds), while not breaking the resonance degeneracies, reduced the range of possible masses somewhat (Figure \ref{fig:hatp7sols}). Finally, only a fraction of the solutions {(about 10\%)} have librating resonant arguments; the ones that do show a much larger amplitude of libration than the reference system ($\Theta_1 \sim 20-40\degr$, $\Delta\varpi \sim 30-70\degr$; see Figure \ref{fig:resonance}). This suggests that the TTV signal alone is not enough to constrain the resonant angles.

Our result is particularly remarkable in that the best-fitting solutions cluster around two different MMRs, preventing a precise characterization of the resonance. Since the two solutions yield a different RV semi-amplitude $K_2$, this degeneracy may be broken with RV observations. Even a small RV dataset, where uncertainty and jitter do not completely wash out the planetary signal, can help constrain the parameters of the perturbing planet to a reasonable level. Indeed, we verified that a second RV dataset comprising 20 measurements with lower jitter ($\sim$ 1 m/s) sufficed to constrain the best-fit solutions to the neighborhood of the 2:1 MMR.  We conclude that while TTVs can be usefully exploited to infer the presence of low-mass perturbing planets, a small number of RV measurements with a precision comparable to $K_2$ is crucial in recognizing the nature of the planetary companion. This fact makes it much more desirable to find configurations orbiting bright parent stars.

\section{HAT-P-13}\label{sec:HATP13}
\begin{figure}
\epsscale{}
\plotone{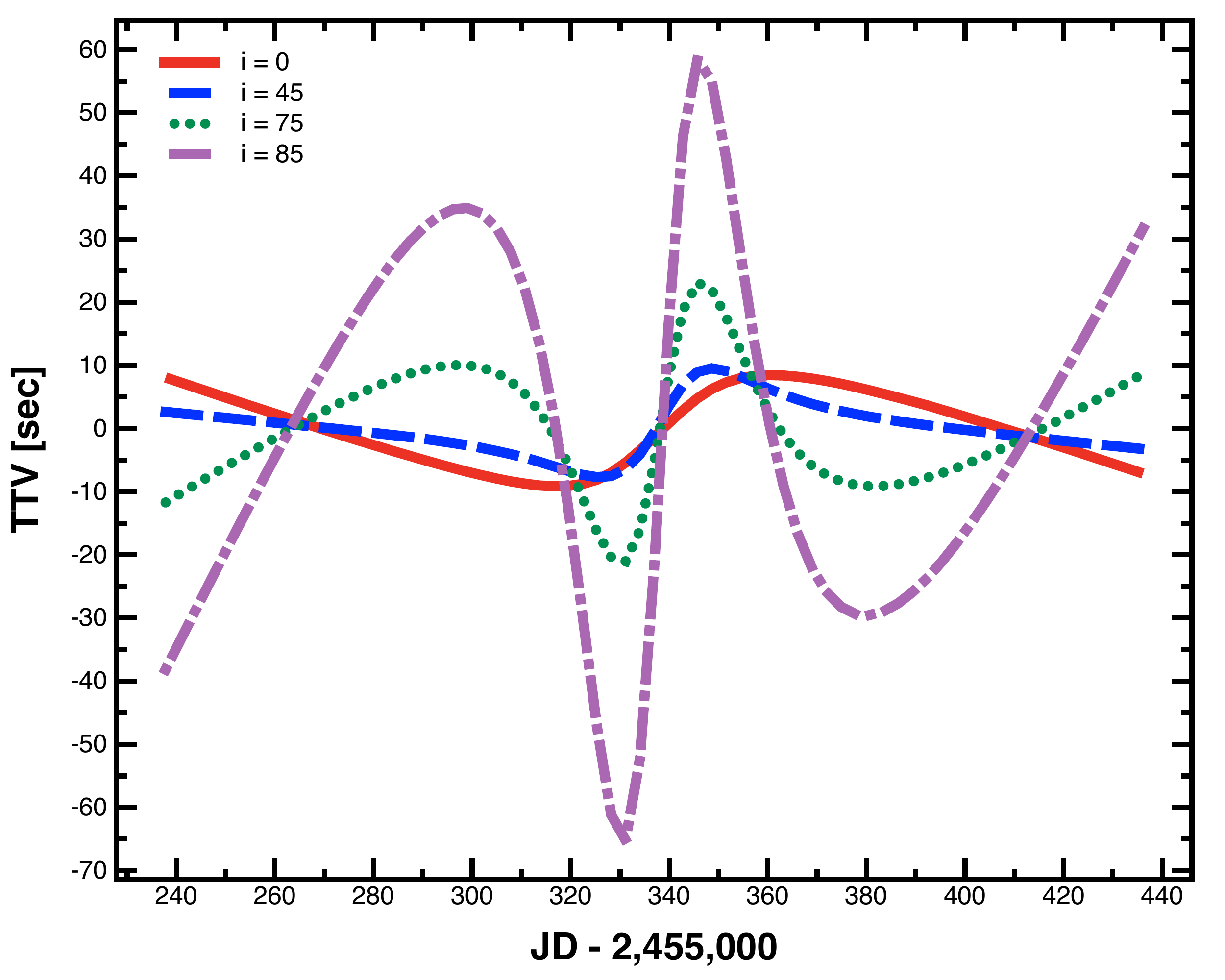}
\caption{Sample TTV signals for four different inclinations of HAT-P-13 c. Orbits close to perpendicular give rise to large TTV signals.}\label{fig:ttv_inc}

\end{figure}
\begin{figure}
\epsscale{0.7}
\plotone{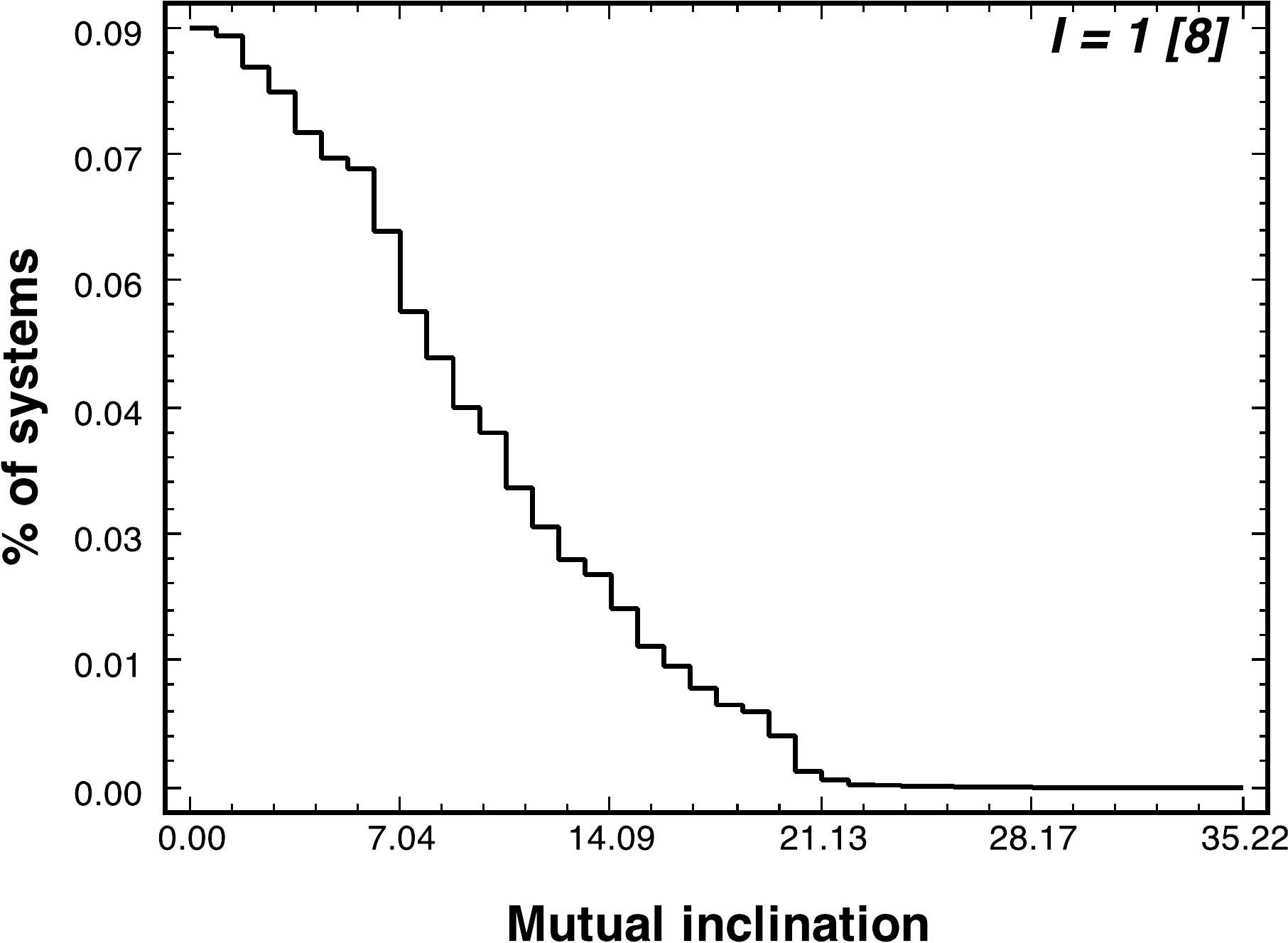}\ 
\plotone{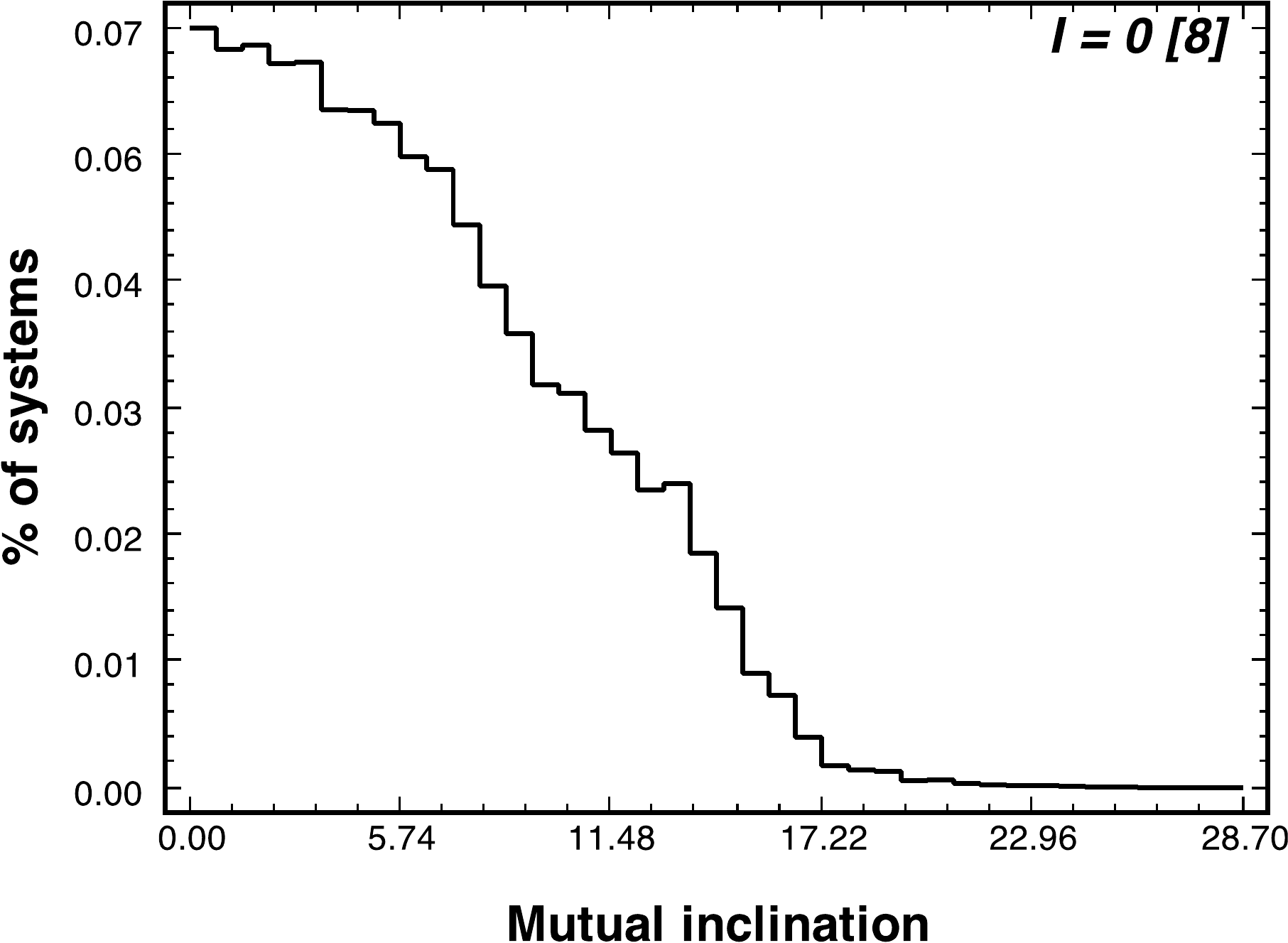}\ 
\plotone{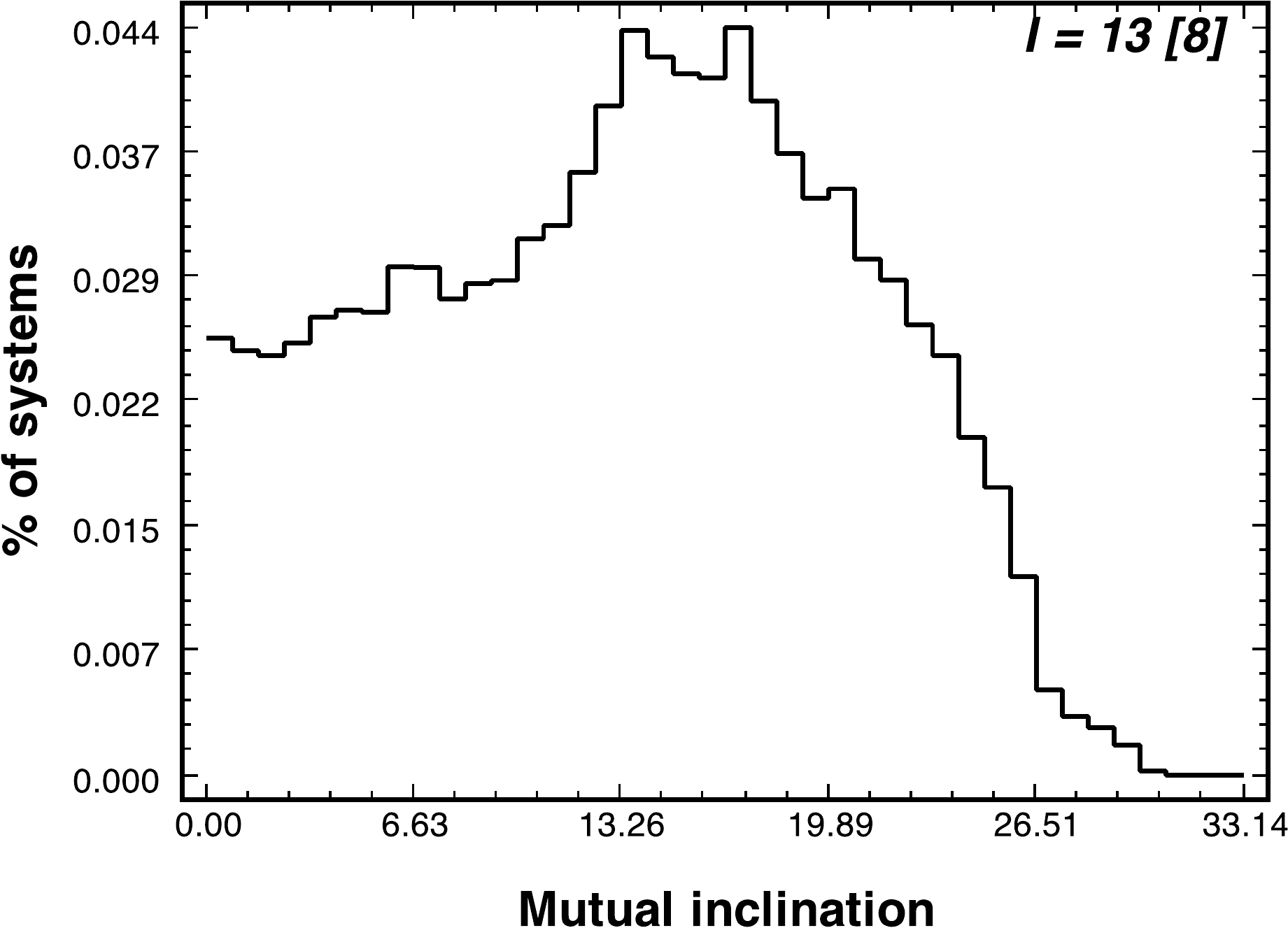}\ 
\plotone{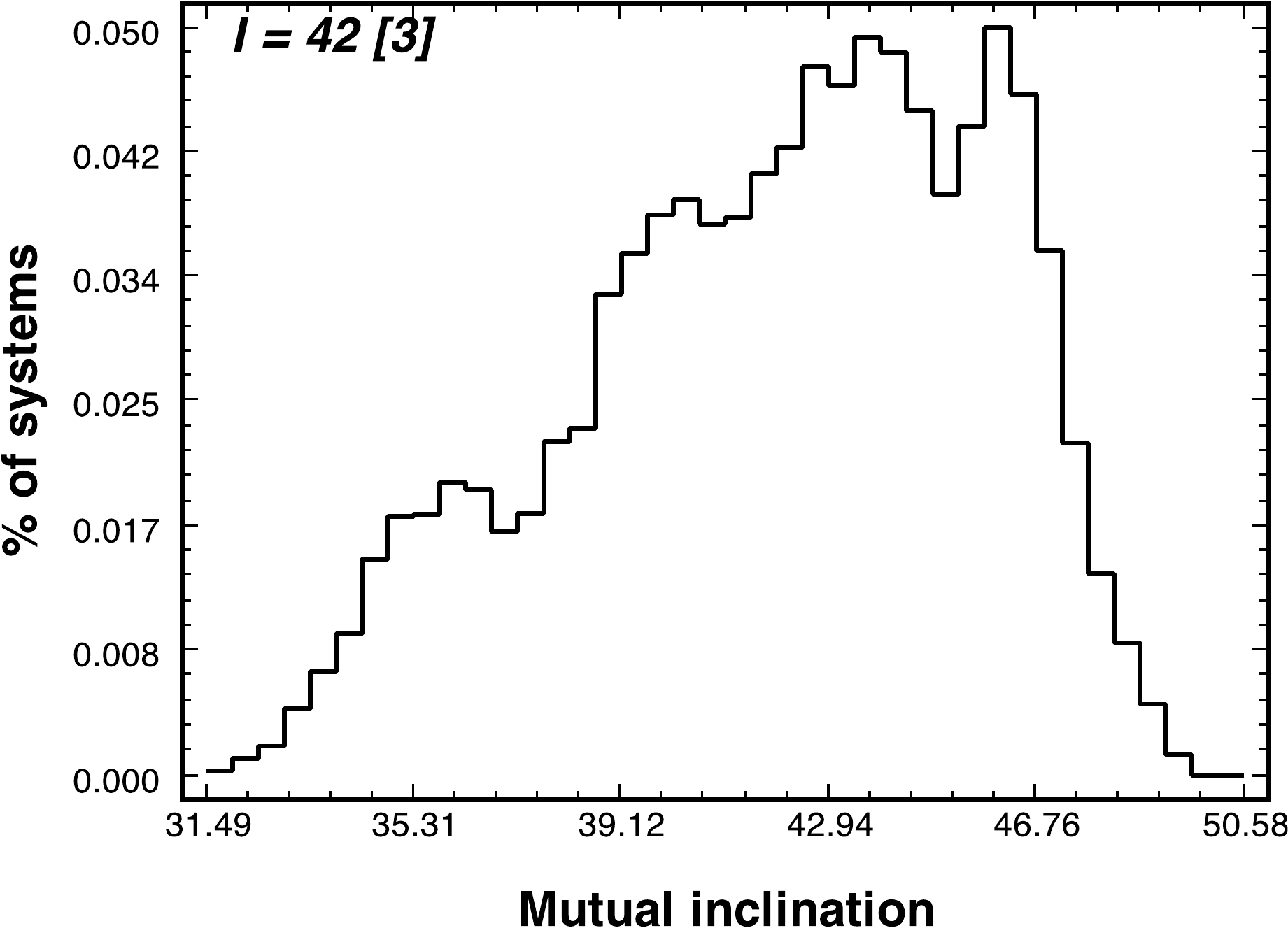}\ 
\plotone{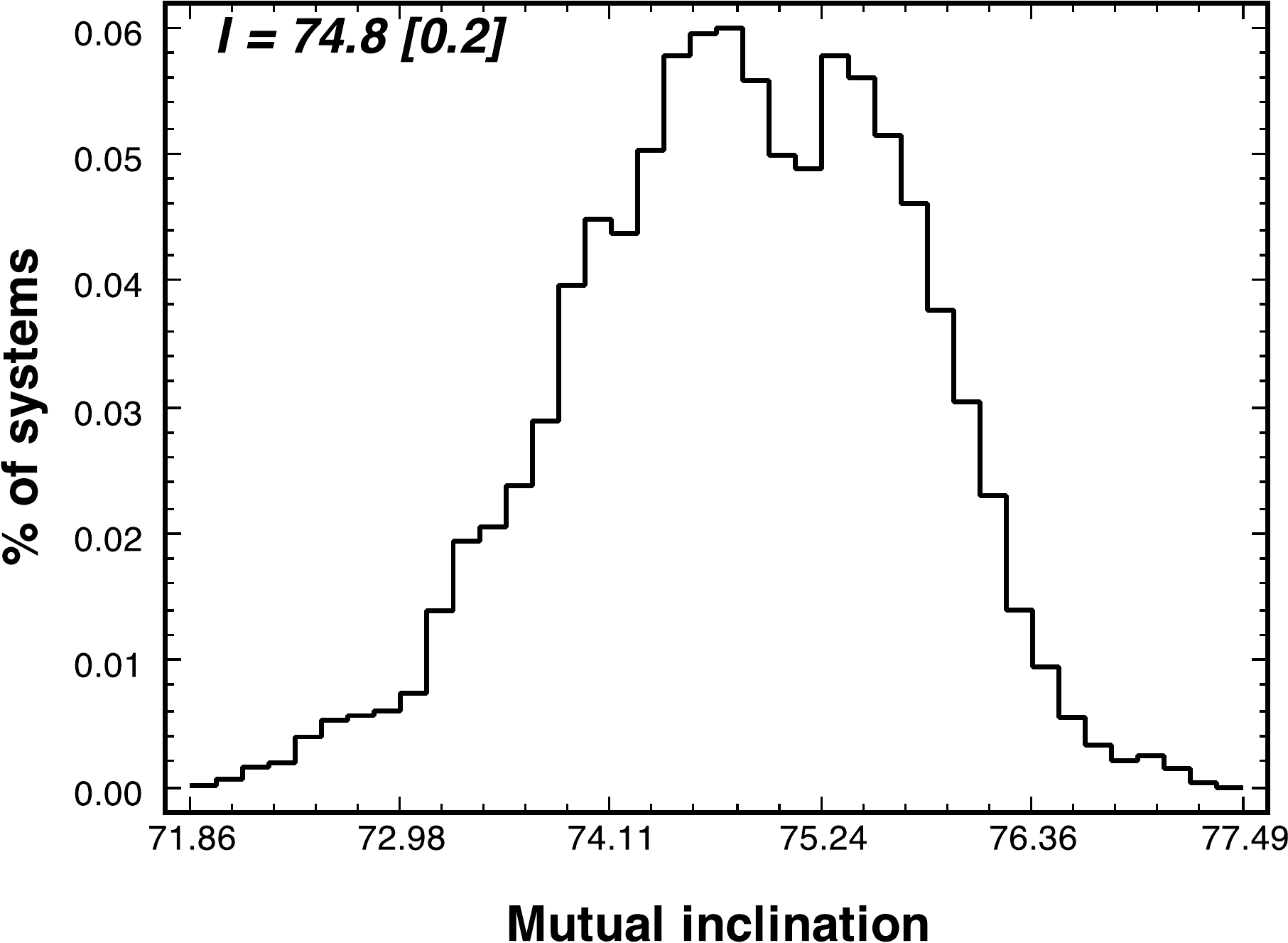}\ 
\caption{Relative inclination distribution for synthetic HAT-P-13 realizations with $I = 0, 5, 15, 45$ and 75\degree{} respectively. The median inclination and standard deviation are given inside each plot.}\label{fig:inc_unc}
\end{figure}

HAT-P-13 was the first system known to contain a transiting planet, \emph{b}, and an eccentric outer planet, \emph{c}, well characterized through RVs \citep{Bakos09}. No transits of planets \emph{c} have been detected thus far. A complete characterization of the three-dimensional configuration of the system can establish the internal structure of planet \emph{b} \citep{Batygin09}  and possibly the formation and scattering history of the system, with certain ranges of inclination being favored on theoretical grounds \citep{Mardling10}. 

Transit timing variations can provide the required constraints on the mutual inclination ($I$) and the nodal line marking the intersection of the two orbital planes ($\Omega$), should transits of \emph{c} not be detected. The amplitude and shape of the TTV signal depend significantly on the two parameters \citep{Payne10}, although this dependence is not trivial. 

Figure \ref{fig:ttv_inc} shows the TTV signal for a number of inclinations.  We centered our dataset around $T_{peri,\ c}$ since the different solutions can be best distinguished by the sharp feature in the neighborhood of the pericenter passage of \emph{c}. While the discovery paper predicted a TTV amplitude of order 15-20 seconds, the updated configuration presented in \citet{Winn10} reduces the expected $\delta t$ near the pericenter passage by a factor $\sim 2$, to about 7 seconds for $I \approx 0$. \citet{Winn10} also measured a prograde Rossiter-McLaughlin effect, suggesting that both orbits are prograde.

We produced several transit datasets for mutual inclinations in the range $0\degree < I < 90\degree$ and $\Omega = 0$, assuming that all transits between $T_{peri,\ c} - 100$ d and $T_{peri, c} + 100$ d are detected; the other elements were drawn using the published uncertainties \citep{Winn10}. We added white noise to the TTV signal at the $4\times 10^{-5}$ d = 3.5 s  level (in order to have $\delta t / \sigma_{tr} > 2$). The RV measurements were generated drawing from the schedule and uncertainties of the Keck/HIRES dataset as reported in the discovery paper.

We used our usual fitting procedure (Bulirsch-Stoer as our integration scheme and Simulated Annealing and AMOEBA in tandem to pinpoint the solution), with the published fit as our starting configuration. When a solution was found, we estimated the uncertainty by running our MCMC algorithm. We generated $4\times 10^6$ trial models; of those, the first 10\% was discarded and only one model every 50 was retained in order to minimize correlations between successive elements of the Markov chain.  Figure \ref{fig:inc_unc} shows the marginal distribution of the fitted relative inclinations for systems with various degrees of inclination ($I = 0\degree, 5\degree, 15\degree, 45\degree$ and $75\degree$). The inclination is well constrained for polar and near-polar configurations of the outer planet, where the TTV signal is sizable; on the other hand, for low inclinations there is a large range of allowed configurations.  However, it is clear that while the inclination distributions are broad, they are consistent with the originating configuration and can discriminate between low-inclination and high-inclination configurations.

\section{Discussion}\label{sec:conc}

In this paper we outlined a procedure to solve the inverse problem of  deriving best-fitting model parameters and associated uncertainties using synthetic radial velocity and transit timing variations datasets simultaneously. The procedure exploits a number of numerical algorithms that are made available to the community through the Systemic Console package. 

We tested our fitting method against a number of synthetic realizations of different planetary configurations, including a system of non-resonating coplanar super-Earths, a system in a deep 2:1  resonance and a non-coplanar system. The transit timing datasets were derived assuming continuous photometric coverage as provided by \Kepler{}, and thus are fully realistic to the extent that the transit timing error can be modeled as white noise with a constant amplitude. Our analysis shows that combined RV and TTV datasets carry enough dynamical information to characterize a system in its full three-dimensional configuration. 

Inverse problems have a storied place in astronomy, with the discovery of Neptune providing a canonical example. In that case, a fortunate orbital geometry allowed Neptune's sky position to be pinpointed with sufficient accuracy that the ``prediction'' of a new planet could credibly be claimed. It is worth pointing out, however, that the accurate ephemeris for Neptune in 1846 was something of a lucky accident. Both Adams' and Le Verrier's masses and semi-major axes were badly off \citep{Grant1852}. The correct position of the planet that emerged from the calculations stems from a degeneracy of solutions during the period surrounding the conjunction of Uranus and Neptune. 

We have found that a similar state of affairs might apply to the transit timing measurement scenarios that will emerge from \Kepler{}. While departure from strict periodicity can be readily measurable, it is generally difficult to work out the complete system configuration from transit timing measurement alone. We confirmed that the suppression of TTV harmonics by the transit timing noise can lead to severe degeneracies in the model parameters, as first pointed out by \citet{Nesvorny08}, even when very low levels of timing error is added to the synthetic data.   In presence of such degenerate set of solutions, however, we have verified that adequate RV data can single out the correct orbital configuration. {We note that other constraints derived by extracting more observables from the photometry, such as the the duration of the transits and their variations \citep[TDV -- ][]{Kipping09, Kipping10}, may also help remove the degeneracies in the solution. Including these contributions will require more sophisticated modelling approaches.}

Finally, we note that our work did not investigate other competing effects that contribute to the TTV signal, chiefly including, but not limited to, light travel time, excitation of tidal modes in the host star, general relativity and the presence of additional planets. Furthermore, the investigation of planetary systems with $N_{pl} >2$ with the methods presented here might be computationally costly due to the large parameter space. 

\acknowledgements
{We acknowledge support from the NSF CAREER Program through grant AST-0449986, and from the NASA Ames Astrobiology Institute through grant NNX08AY38A.}

\bibliographystyle{apj}

\end{document}